\documentclass[twocolumn,
pra
,superscriptaddress
,floatfix
,aps
,10pt
,showpacs]{revtex4-2}
\usepackage[dvipsnames]{xcolor}
\usepackage{amsmath,amsfonts}
\usepackage{graphicx}
\usepackage{makecell}

\begin{document}
\title{Spontaneous coherence in spatially extended photonic systems: Non-Equilibrium Bose-Einstein condensation 
}

\author{Jacqueline Bloch}
\affiliation{Universit\'e Paris-Saclay, CNRS, Centre de Nanosciences et de Nanotechnologies, 91120, Palaiseau, France}

\author{Iacopo Carusotto}
\affiliation{INO-CNR BEC Center and Dipartimento di Fisica, Università di Trento, 38123 Trento, Italy}
\author{Michiel Wouters}
\affiliation{TQC, Universiteit Antwerpen, Universiteitsplein 1, B-2610 Antwerpen, Belgium}

\date{\today}

\begin{abstract}
In this review, we give an interdisciplinary overview of Bose-Einstein condensation phenomena in photonic systems. We cover a wide range of systems, from lasers to photon condensates in dye-filled cavities, to excitons in semiconductor heterostructures, to microcavity polaritons, as well as emerging systems such as mode-locked lasers and classical light waves. Rather than diving into the specific properties of each system, our main focus will be to highlight those novel universal phenomena that stem from the driven-dissipative, non-equilibrium nature of these systems and affect the static, dynamic and coherence properties of the condensate. We conclude with our view on the future perspectives of this field for both fundamental science and technological applications.
\end{abstract}

\maketitle

\section{Introduction}


Since its original prediction in 1924, Bose-Einstein condensation has been a major focus of interest for theoretical and experimental research~\cite{Huang,BECbook}. In parallel to the impressive achievements in gases of ultracold atoms, 
exciting new phenomena have been uncovered by the study of condensation effects in photonic systems.

Whereas the same spontaneous $U(1)$ symmetry breaking mechanism underlies standard textbook BEC and the onset of spontaneous coherence in photonic systems~\cite{Graham:ZPhys1970,DeGiorgio:PRA1970}, the unavoidable photon losses make these latter to be intrinsically non conservative systems. As a result, their steady-state departs from the usual thermodynamical equilibrium state and is rather determined by a dynamical equilibrium of driving and dissipation~\cite{schmittmann1995statistical,Cross:RMP1993,carusotto:2013}. As a consequence, condensation displays radically new features in both its static, 
dynamical, 
superfluid, 
and fluctuation 
properties. 

In this review, our goal is to provide a global picture of the field of non-equilibrium condensation in photonic systems and give a comparative review of the main photonic platforms where landmark experiments have been performed, such as lasers and nonlinear optical devices~\cite{lundeberg2007spatial,staliunas2002transverse,krupa2017spatial,baudin2020rayleigh,oren2014classical}, photon~\cite{klaers2010bose}, excitons~\cite{High:Nature2012,Alloing2014} and polariton~\cite{Stevenson:PRL2000,Deng:Science02,Kasprzak:Nature2006,sun2017bose} condensates. In doing this, our main focus will be on the identification of those universal phenomena that stem from the non-equilibrium condition rather than from the specific features of each system. Our interdisciplinary point of view involves fundamental concepts from quantum optics, non-equilibrium statistical mechanics, quantum condensed matter and quantum field theories, and opens promising perspectives towards new developments in fundamental science as well as towards applications in optoelectronics~\cite{colombelli2015perspectives,Sanvitto2016,fraser2016physics,BallariniDeLiberato,SegevBandres2021, colombelli2015perspectives}, 
analog computation~\cite{berloff2017realizing,Davidson_laser2013,mcmahon2016fully}, and quantum technologies~\cite{ma2019dissipatively,Roushan:2016NatPhys,clark2020observation}.

Our review is organized as follows. Sec.\ref{sec:BEC} will present a general definition of condensation in terms of long-range correlations in the photon field, which 
applies equally well to equilibrium and non-equilibrium systems. In Sec.\ref{sec:platf}, we will provide a comparative review of the main different experimental platforms and of their regimes of operation. 
In Sec.\ref{sec:non_eq}, we will review the new phenomena that have been predicted to arise from the non-equilibrium nature of the condensation process and we will present their most celebrated experimental observations. In Sec.\ref{sec:conclu}, we present our conclusions on the state of the field and we outline our view on its perspectives for both fundamental and applied sciences.

\section{What is condensation?}
\label{sec:BEC}

The phenomenon of Bose-Einstein condensation (BEC) was originally predicted
in the context of the quantum statistical mechanics of an ideal Bose gas~\cite{Huang,BECbook}. Its most celebrated signature consists of the accumulation of a macroscopic fraction of the particles into the lowest energy single-particle state. While first experimental hints for a finite condensed fraction were obtained by neutron scattering off superfluid $^4$He~\cite{BECbook}, definitive evidence came from the observation of the $\mathbf{k}=0$ peak in the momentum distribution of ultra-cold atomic gases~\cite{Anderson:Science1995,Davis:PRL1995}.

\begin{figure*}[htbp]
    \centering
    \includegraphics[width=\textwidth]{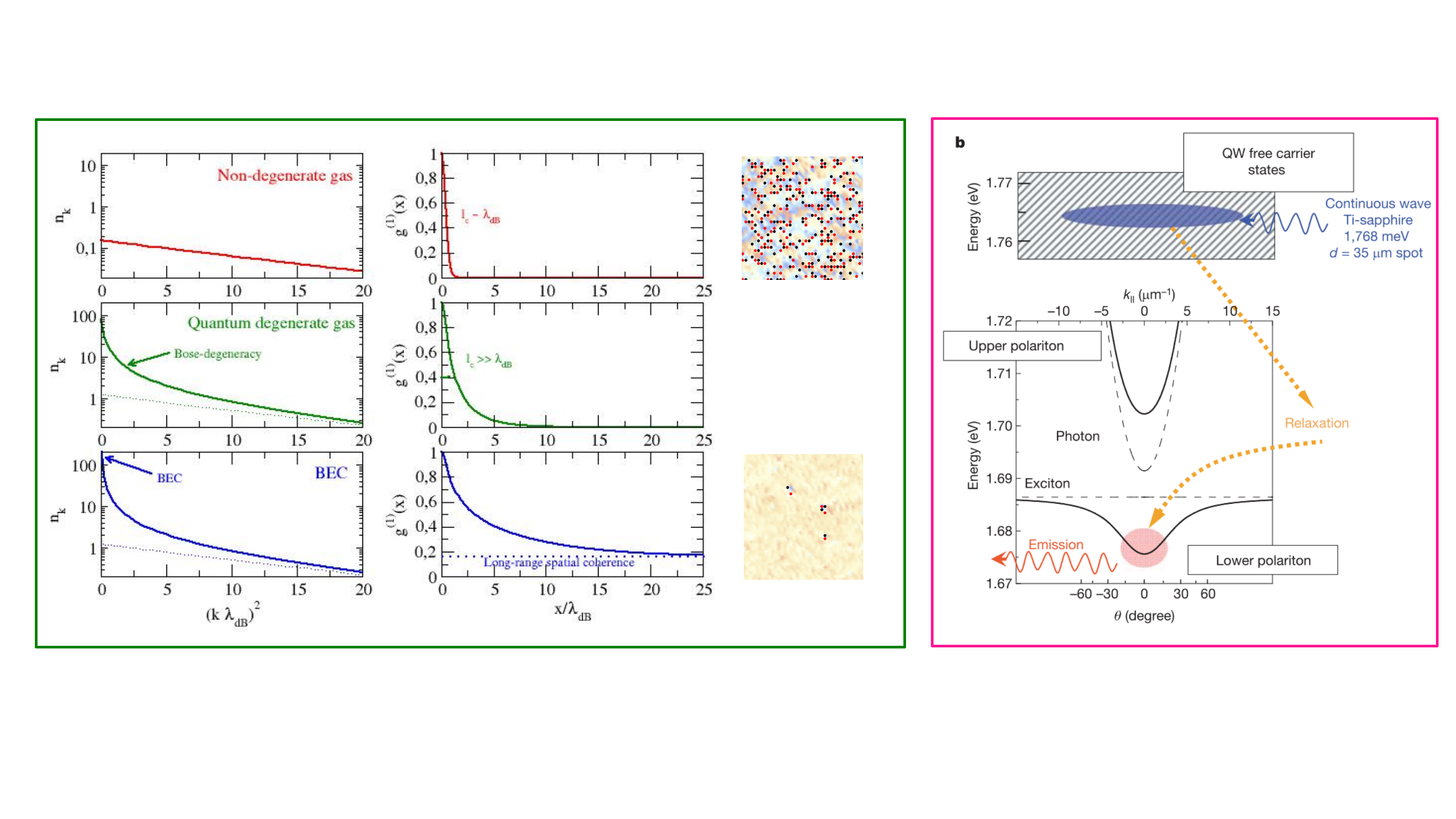}
    \caption{General features of condensation. Left frame: momentum distribution (left) and first-order spatial coherence (right) of a Bose gas with a fixed temperature and a growing density, ranging from non-degenerate gas with a Maxwell-Boltzmann distribution (top), to a quantum degenerate gas (middle), to a Bose-condensed gas with long-range spatial coherence (bottom). Insets show examples of the density (background color) and vortex (red and black dots) distribution for a randomly chosen realization of the matter field at the given average density and temperature. Insets from Ref.\onlinecite{dagvadorj2015nonequilibrium}. Right frame: sketch of the pump and loss mechanisms at play in an exciton-polariton condensate. Figure from Ref.\onlinecite{Kasprzak:Nature2006}.}
    \label{fig:BEC}
\end{figure*}

A deeper insight in the BEC transition can be obtained from its interpretation as a second-order phase transition spontaneously breaking the $U(1)$ symmetry associated to the phase of a complex-valued matter field~\cite{Gunton:PR1968,CCT:CdF}. This description highlights the close analogy with other well-known phase transitions such as ferromagnetism~\cite{ferro_book} and provides a clear physical picture of the underlying phase transition.  

In the disordered phase at high temperatures, the spins of a ferromagnet have a random orientation with a microscopically short coherence length. Analogously, as it is sketched in the top row of the left frame of Fig.\ref{fig:BEC}, the quantum matter field $\hat\Psi(\mathbf{r})$ at high temperature/low density consists of a superposition of incoherent waves whose coherence length is set by the thermal De Broglie wavelength $\lambda_{dB}=\sqrt{2\pi\hbar^2/mk_B T}$. 
As the temperature is lowered (or the density increased), quantum degeneracy effects due to the Bose statistics of the particles make the coherence length of the matter field to grow (middle row) and eventually diverge as the BEC critical point is approached. 

Below the critical temperature, a ferromagnet is characterized by a uniform magnetization throughout the whole sample. 
Analogously, at low temperatures (high densities) a condensate appears in the Bose gas, characterized (bottom row) by a long-range spatial coherence in the matter field,
\begin{equation}
    \lim_{|\mathbf{r}-\mathbf{r}'|\to \infty} \left\langle \hat\Psi^\dagger(\mathbf{r})\,\hat\Psi(\mathbf{r}')\right\rangle = \phi_o^*(\mathbf{r})\,\phi_o(\mathbf{r}')\neq 0,
    \label{eq:g1_BEC}
\end{equation}
where the complex-valued classical order parameter $\phi_o(\mathbf{r})$ physically corresponds to the wavefunction of the macroscopically occupied single-particle state. 
This long-range coherence is independent of the nature of the bosonic field: it is at the heart of the matter-wave interference experiments that offered an ultimate evidence of condensation in ultracold atomic gases~\cite{andrews1997observation}, and underlies the use of photonic condensation effects as a source of coherent light.


In contrast to the extremely long, virtually infinite lifetime of typical condensed-matter systems such as liquid Helium or ultracold atomic gases, a general property of optical systems is the (almost) unavoidable presence of significant particle losses, resulting both from absorption and from the radiative decay experienced by photons in confined geometries. As a result, some external pump is needed to compensate for losses and sustain a non-vanishing light intensity. 
Even under a continuous-wave pump, the steady-state that the system attains  is typically distinct from a standard thermodynamical equilibrium state and is rather determined by a dynamical balance of pumping and dissipation. 
While the macroscopic wavefunction $\phi_o$ of a zero-temperature atomic condensate is ruled by the celebrated Gross-Pitaevskii equation (GPE)~\cite{BECbook} analogous to the Landau-Lifshits equation for the magnetization dynamics in ferromagnets, generalized forms of GPE including additional terms for driving and dissipation are required for the description of photonic condensates and give rise to a way richer dynamics~\cite{carusotto:2013}.

Rather than being just a hindrance, the presence of radiative decay channels in optical systems is a crucial asset for experimental investigations: in this way, all coherence properties of the condensate get transferred to the radiated field and can be retrieved with standard quantum optical tools. In particular, condensation can be assessed via the definition in Eq.\ref{eq:g1_BEC} by looking at the large-distance decay of the first-order coherence function \begin{equation}
  g^{(1)}(\mathbf{r},\mathbf{r}')=\frac{\left\langle \hat\Psi^\dagger(\mathbf{r})\,\hat\Psi(\mathbf{r}')\right\rangle}
  {\left[\left\langle \hat\Psi^\dagger(\mathbf{r})\,\hat\Psi(\mathbf{r})\right\rangle
  \left\langle \hat\Psi^\dagger(\mathbf{r}')\,\hat\Psi(\mathbf{r}')\right\rangle \right]^{1/2} }
\end{equation} 
of the emitted light.

Still, as it was first realized in~\cite{Graham:ZPhys1970}, the long-range order of a photonic condensate is unavoidably spoiled by quantum fluctuations in the $d\leq 2$ dimensionalities that are currently used in experiments. Quite remarkably, this prediction appeared shortly after related work in the context of equilibrium statistical mechanics, in particular liquid Helium~\cite{Mermin:PRL1966,Hohenberg:PR1967,Reatto:PR1967}. These fluctuations turn the condensate into a quasi-condensate~\cite{BECbook}, whose coherence slowly decays at large distances similarly to the equilibrium case~\cite{Wouters:PRB2006,Szymanska:PRL2006,dagvadorj2015nonequilibrium} but also displays peculiar spatio-temporal properties related to its non-equilibrium condition~\cite{altman2015two,wachtel2016electrodynamic,gladilin2014spatial,He:PRB2015,ji2015temporal,squizzato2018kardar,zamora2017tuning,he2017space,Fontaine21}.  

There is no doubt that laser operation is the most celebrated example of non-equilibrium condensation phenomenon in the optical context. Upon crossing the laser threshold, the statistics of the emitted light transforms from an incoherent thermal state to a coherent state with a well-defined electric field amplitude and phase. Even though a description of laser operation in terms of a phase transition has been usefully adopted to describe the dynamics of generic devices since the early days of laser physics~\cite{Graham:ZPhys1970,DeGiorgio:PRA1970}, strictly speaking it is only valid in spatially extended systems where a continuum of cavity modes is available for lasing and it is meaningful to consider the long-distance correlation functions involved in the condensation criterion (\ref{eq:g1_BEC}).

A most relevant class of devices for this purpose are the so-called Vertical Cavity Surface Emitting Lasers (VCSELs)~\cite{wilmsen2001vertical,Iga_2008}, which can be built with arbitrarily wide lasing areas. For a series of reasons, typically related to different sources of instabilities that easily makes the condensate to split into a multi-mode emission pattern (see, e.g., Refs. \onlinecite{HegartyPRL1999,lugiato2015nonlinear,longhi2018invited}, and references therein), these systems have not yet been fully exploited for studies of the basic physics of condensation phenomena.


\begin{figure*}[htbp]
    \centering
    \includegraphics[width=\textwidth]{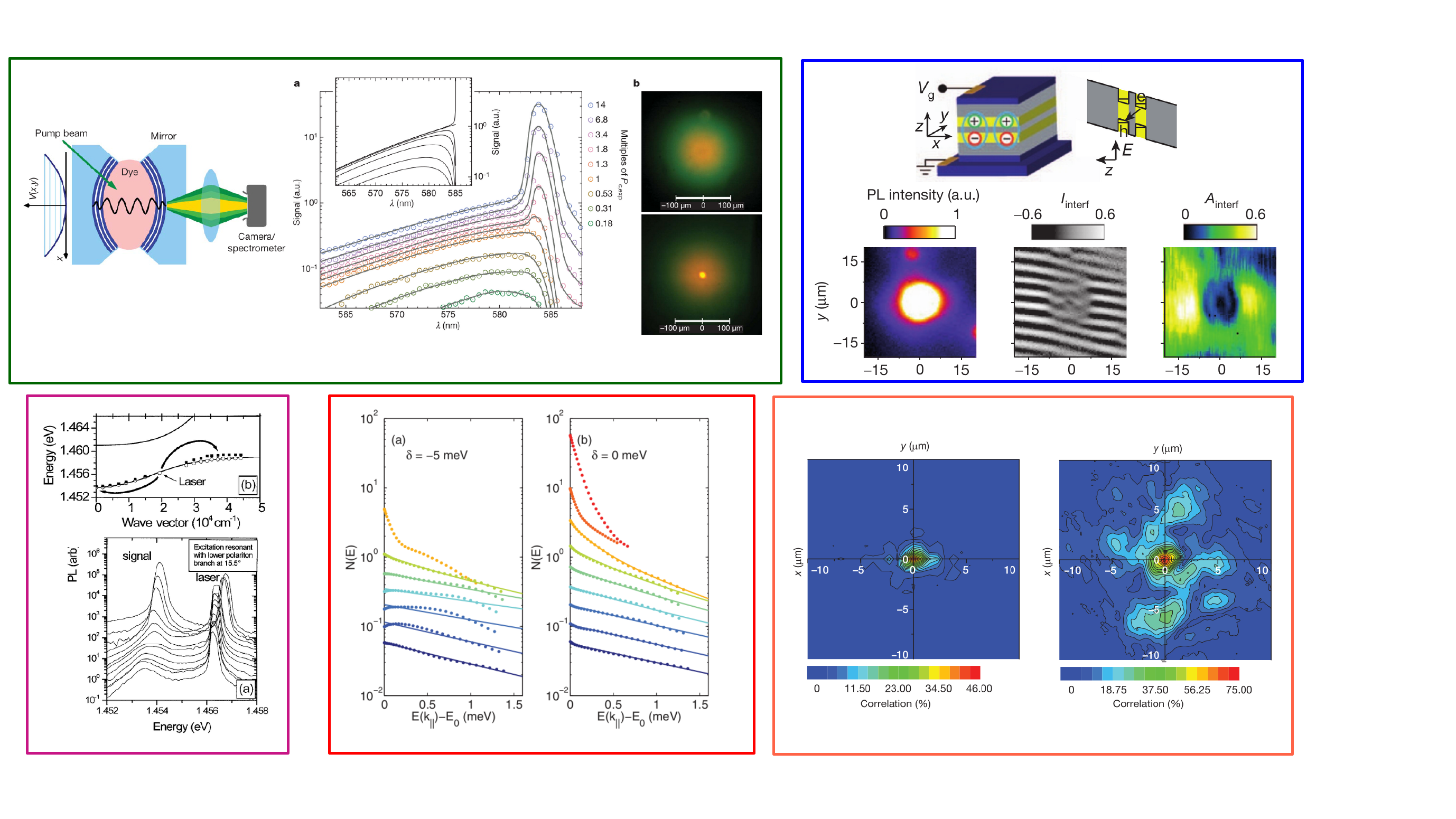}
    \caption{BEC of photons, excitons and exciton-polaritons. Top-left frame: BEC of photons. Scheme of the set-up (left), wavelength distribution of the emission for different pump strengths, showing the appearance of a BEC peak at long wavelengths for strong enough pump intensity (center), spatial profile of the emission for weak (top-right) and strong (bottom-right) pump intensities. Figure from Ref.\onlinecite{klaers2010bose}. Top-right frame: BEC of excitons.  Sketch of the structure of the indirect excitons considered in the experiment (top). The condensation signatures are visible in the intensity (bottom-left), interferogram (bottom-center) and coherence (bottom-right) of the emission around a localized bright spot. Figure from Ref.\onlinecite{High:Nature2012}. Bottom-left frame: OPO condensation of polaritons. Scheme of the polariton parametric scattering  process (top). Energy-distribution of the emission for increasing pump strengths, showing the appearance of a coherent signal peak for strong enough pump power. Figure from Ref.\onlinecite{Stevenson:PRL2000}. Bottom-center frame: incoherently pumped polariton BEC. Energy-distribution of the emission for increasing pump strength, showing the appearance of the BEC peak at low energies for strong pump power. The two panels refer to different values of the cavity-exciton detuning in respectively the non-thermalized (left) and thermalized (right) regimes.  Figure from Ref.\onlinecite{sun2017bose}. Bottom-right frame: Spatial map of the coherence of an incoherently pumped polariton gas for two values of the pump strength, in either the non-degenerate (left) and Bose-condensed (right) regimes. Figure from Ref.\onlinecite{Kasprzak:Nature2006}. }
    \label{fig:cavity_BEC}
\end{figure*}

\section{Landmark experiments}
\label{sec:platf}


Besides lasers, in the last decades a number of other optical systems have been employed to investigate condensation phenomena in different regimes. While the basic symmetry breaking mechanism is the same in all these condensation phenomena, the properties of these systems smoothly connect the strongly non-equilibrium laser operation to equilibrium BEC passing through a sequence of configurations with growing values of the collisional thermalization to loss ratio and different values of the inter-particle interaction constant. The interplay of these phenomena is at the heart of a rich variety of different phenomenologies: this Section will be devoted to a survey of a few most interesting such systems.

\subsection{Photon condensates}

A class of systems that received a strong attention as an example of BEC of photons consists of light  in optical cavities filled by a dye solution at room temperature~\cite{klaers2010bose}. In this class of systems, illustrated in the top-left frame of Fig.\ref{fig:cavity_BEC}, the photons are in a thermal equilibrium condition at a relatively high temperature, so the macroscopically occupied photon mode is supplemented by a substantial thermal component which accurately follows the textbook Bose-Einstein distribution~ \cite{klaers2010bose,greveling2018density}. 
However, in contrast to the case of isolated Bose gases, e.g. of cold atoms, the thermal distribution does not originate from collisions between photons in the gas but rather from a thermalization mechanism between the photons and the excitations in the dye. For a dye at temperature $T$, the ratio of emission to absorption coefficients satisfies a  Kennard-Stepanov relation $R_{\rm em}(\omega)/R_{\rm abs}(\omega) \sim e^{-\hbar\,\omega/k_BT}$~\cite{kennard,stepanov,moroshkin}.
It is this condition which imprints the thermal distribution on the photon gas. Of course, this mechanism is effective only as long as the dye absorption/emission rate is much larger than the cavity losses. For stronger losses, thermal equilibrium is instead lost and the system recovers the behaviour of a regular laser~\cite{schmitt.PhysRevA.92.011602,walker2018driven,Hesten:PRL2018}. 
Interestingly, a similar thermal tail in the photon distribution has also been observed in a semiconductor VCSEL device~\cite{Bajoni:PRB2007}, where it was interpreted following semiconductor laser textbooks~\cite{chuang2012physics} as the result of spontaneous emission from the radiative recombination of a thermalized uncorrelated electron/hole gas. Interesting observations of thermalization and condensation effects in a semiconductor resonator in the weak-coupling regime have recently appeared in Ref.\cite{Barland:21}.

The most elementary theoretical model to describe photon condensates consists of rate equations for the number of photons in the various trap levels and excited molecules \cite{schmitt2018}. A more sophisticated quantum optical description was developed in Ref.~\cite{kirton-keeling-13}, showing that thermal equilibrium is lost when cavity losses become of the order of the absorption rate and the system then behaves more like a regular laser.
In regimes where all relevant modes are largely occupied, computationally less costly stochastic classical field approaches can also be used~\cite{gladilin2020classical}.

\begin{figure*}[htbp]
    \centering
    \includegraphics[width=\textwidth]{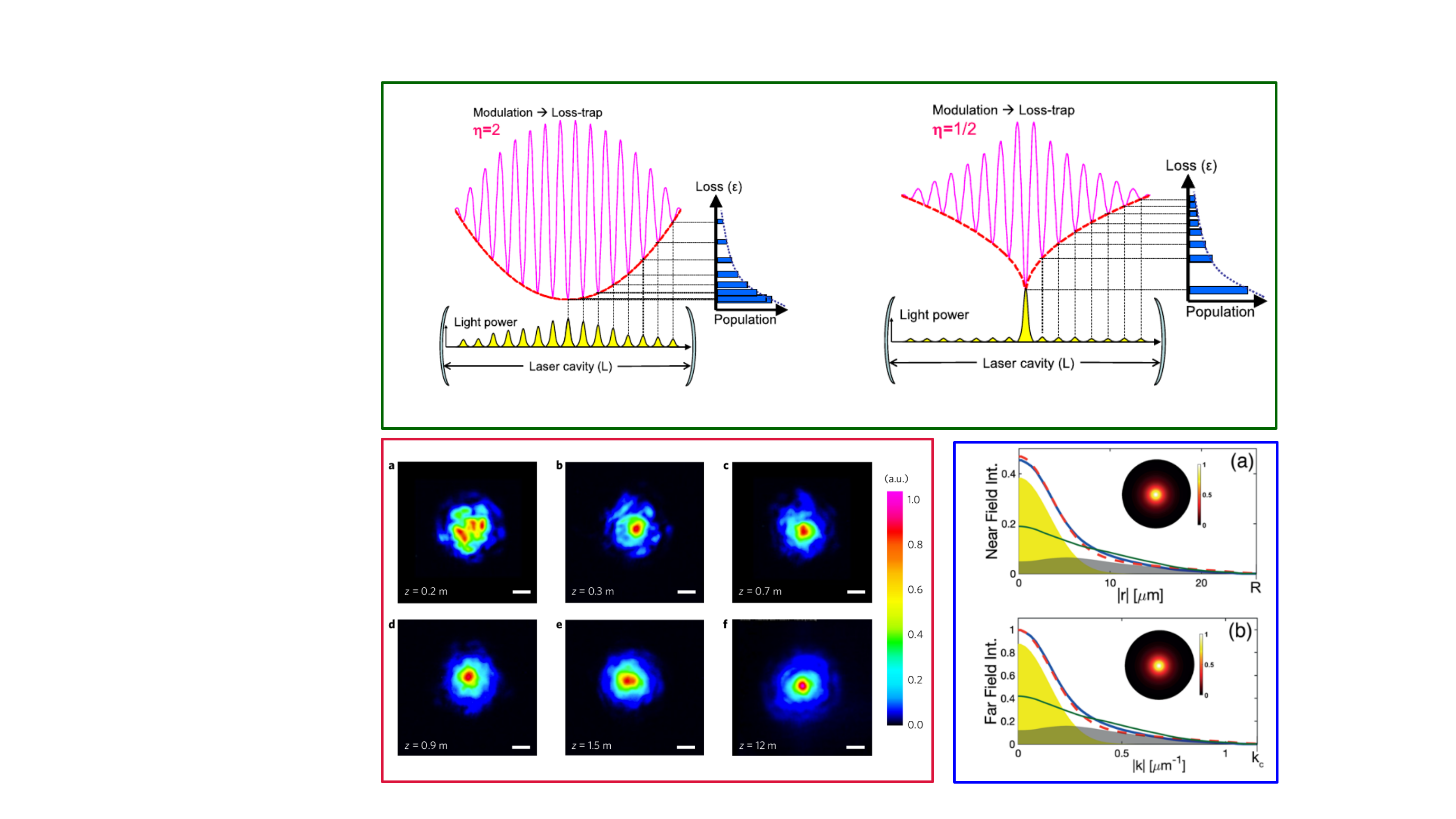}
    \caption{Novel platforms for condensation. Top frame: photon condensation in a mode-locked erbium-doped fiber laser. Here, the role of the different modes of a cavity is played by the different pulses (yellow areas), whose relative losses can be controlled via the details of the mode-locking modulation (red lines). 
    While for a large $\eta=2$ loss exponent the intensity (blue histograms) smoothly distributes among many pulses (left), for small $\eta=1/2$ condensation of the intensity into a single pulse is observed (right). Figure from Ref.\onlinecite{oren2014classical}. Bottom frames: thermalization and condensation of classical light waves in nonlinear graded-index waveguides. Left: self-cleaning of the spatial profile as light propagates along the waveguide. Figure from Ref.\onlinecite{krupa2017spatial}. Right: near-field (top) and far-field (bottom) profiles of the thermalized beam at the output of the waveguide, displaying a sizably occupied condensate. The dashed-red lines are the theoretical predictions of the Rayleigh-Jeans distribution, to be compared with the experimental measurements shown as blue lines. The green lines show the non-thermal distributions in the incident field at the entrance of the waveguide. Figure from Ref.\onlinecite{baudin2020rayleigh}. 
    }
    \label{fig:other_plat}
\end{figure*}

The main ingredient needed to achieve photon condensation, namely an energy dependent ratio of gain to losses, can be achieved not only by means of a thermalized gain medium, but also by directly engineering the photon losses. This approach was followed in~\cite{oren2014classical}, where  a suitably designed active mode-locking mechanism in an erbium-doped fiber laser enabled engineering of a many-light-pulse system experiencing an effective loss trap.
The flexibility of this platform allowed to engineer the spatial dependence of the loss rate in the form $\gamma(x) \sim |x|^\eta$, where the effective spatial coordinate $x$ indicates the different pulses and $\eta$ is a tunable loss exponent.
Examples of the observed distribution of light intensity over the different pulses  are shown in the top panel of Fig.\ref{fig:other_plat} for different values of the loss exponent $\eta$: for low enough noise level and sufficiently small $\eta$, condensation is visible as a concentration of the laser emission into the lowest loss pulse.

\subsection{Exciton condensates}

Whereas photon condensates are based on photons weakly coupled to an optically active material, the opposite limit of bosonic quasi-particles directly emerging from matter excitations has also attracted a strong interest for many decades. Such a research direction started with the pioneering predictions of exciton BEC in the early '60s~\cite{blatt1962bose,moskalenko1962inverse,keldysh1965possible}:
since excitons are made of electron-hole pairs bound by Coulomb interaction, they present a very light effective mass as compared to standard material particles like atoms. As a result, the critical temperature for exciton BEC was expected to be around 1K for realistic exciton densities and thus accessible with standard cryogenic techniques. 

The main issue for the onset of exciton BEC is the difficulty  for optically generated excitons to  thermalize within their lifetime. 
Among all semiconductor materials displaying excitons, ${\rm Cu}_2{\rm O}$ seemed very promising in this respect since radiative recombination is forbidden by crystal symmetry and the recombination time can reach tens of ms. In spite of the intense research effort devoted to exciton BEC in ${\rm Cu}_2{\rm O}$ in the last 40 years, no clear success was reported. The main obstacle appears to be Auger recombination, see~\cite{Snoke2014} for a dedicated review. 

A crucial advance to circumvent these difficulties was made possible by the use of indirect excitons in semiconductor heterostructures, which involve spatially separated electrons and holes in either spatially wide quantum wells under a strong electric field or in double quantum wells (top panel of the top-right frame of Fig.\ref{fig:cavity_BEC}). Also in this case, the weak spatial overlap between the electron and hole wavefunctions results in long exciton lifetimes (up to tens of ms at low temperature) with the additional advantage that exciton thermalization times can be very short (typically in the ns range below 1 K in GaAs based heterostructures~\cite{IvanovPRB99}). First experimental indications for BEC of indirect excitons were reported in 2002 in a geometry where hot excitons travel away from a localized excitation region and form a large bright ring where the condensate emerges~\cite{Butov2002,Snoke2002}. A decade later, extended spatial coherence in the region close to the exciton ring  was finally demonstrated by interferometrical methods, offering a direct signature for exciton BEC~\cite{High:Nature2012}, as illustrated in the bottom panels of the top-right frame of Fig.\ref{fig:cavity_BEC}. 

In spite of this evidence, quantitative analysis of these observations appeared to be delicate. Indeed, in GaAs quantum wells, the exciton ground state is actually a dark $J=2$ state~\cite{Combescot07}, which could result in a dark condensate, not accessible to optical techniques. Further investigation however showed that fermion exchange and collision between dark excitons \cite{Combescot2012,Shiau19} provide mechanisms for the recovery of a small bright component for the exciton condensate, which is then named a ``grey'' condensate. As a result, exciton condensates should be localized in low emission intensity regions where the density has to be estimated from the renormalization of the recombination energy by exciton-exciton interaction \cite{Voros09,Shilo2013, Combescot2014}. This picture was firmly confirmed in later works~\cite{Alloing2014,Combescot2017}, which demonstrated that the highest spatial coherence is indeed observed in region of maximum exciton density and small brightness and thus provided a convincing evidence of exciton condensation in semiconductor heterostructures. Sophisticated quantum Monte Carlo studies of exciton condensates were reported in~\cite{lozovik2008bose}, hinting at the strongly-correlated nature of these systems.

In the last years, a new platform for the search of exciton condensates at higher temperatures is being provided by 2D materials~\cite{Wang2019}. Parallel to these advances, many other systems have also been intensely studied in the context of BEC of quasi-particles in solid-state systems, including excitons in doped bilayers~\cite{eisenstein2014exciton} and  magnons~\cite{Demokritov:Nature2006,Giamarchi:Nature2008}. In spite of their great interest, all these systems go beyond the scope of our review and will not be discussed further here. 

\subsection{Exciton-polariton condensates}
Exciton-polaritons are a combination of the two  systems considered in the previous Sections. They are hybrid exciton-photon bosonic quasi-particles which emerge from the strong coupling between direct excitons confined in quantum wells and photons confined in a cavity \cite{kavokin2017microcavities}. In planar microcavities, they form two bands, the so-called upper and lower polariton~\cite{Weisbuch:PRL1992}. The dispersion of the lower polariton band features a characteristic s-like shape and inherits from the photonic component a very light effective mass close to the center of the Brillouin zone, typically around $10^{-5}$ the free electron mass~\cite{HoudrePRL1994}. Thanks to this light effective mass, polariton condensation was anticipated to occur at higher temperatures than exciton BEC, namely above 10 K and possibly up to room temperature. 

The originally proposed scheme~\cite{Imamoglu:PRA1996} was based on the injection of high-energy electron-hole pairs by a non-resonant optical excitation beam, as sketched in the right frame of Fig.\ref{fig:BEC}. 
The relaxation mechanism considered for thermalization relied on stimulated exciton-phonon interaction. Even though a bottleneck effect in the phonon relaxation process turned out to be a serious obstacle against polariton thermalization 
\cite{Tassone97}, an efficient alternative relaxation channel toward the bottom of the lower branch is provided by exciton-exciton scattering~\cite{Tassone:PRB1999}. 
Since these mechanisms operate in the excitonic regime well below the Mott density, the coherent light emission does not require the population inversion that is instead needed in standard lasing. This is most promising for the development of coherent light sources with very low threshold.

Less than ten years after its proposal, first experimental hints of polariton condensation were reported in CdTe based samples~\cite{Dang:PRL1998} and, then, in GaAs based cavities~\cite{Deng:Science2002,Balili:Science2007}: above some threshold value of the excitation intensity, a strong non-linear emission was observed together with the onset of temporal coherence as measured via either a spectral narrowing of the emission or a reduction of the intensity fluctuations. As a final evidence of BEC, extended spatial coherence was finally demonstrated in a CdTe cavity~\cite{Kasprzak:Nature2006} as shown in the bottom-right frame of Fig.\ref{fig:cavity_BEC}. 

In the following years, many groups investigated these polariton BEC in details~\cite{Deng:RMP2010,Byrnes2014}. A key issue that was fiercely debated in the community was to clearly demonstrate that the coherent emission was indeed coming from a polariton mode with mixed light-matter character and not from a photon mode, as it instead happens in a regular laser. Following the shift of the emission energy for growing pump intensity allowed to distinguish the two regimes~\cite{KiraPRL1997,Butte:PRB2003,Deng:PNAS2003}, as well as to observe a second threshold for photon lasing~\cite{BajoniPRL2008}. The fine details of the intermediate regimes between strong and weak coupling are still the subject of intense work~\cite{PhysRevX.11.011018}.

Another excitation scheme to generate a polariton condensate is based on parametric scattering. Because of their excitonic component, polaritons inherit a strong Kerr non-linearity. By resonantly exciting the system close to the inflection point of the lower polariton branch, it is possible to trigger a triply-resonant parametric scattering from two polaritons at the pump wavevector into one polariton in a signal mode close to $k=0$ and one polariton in an idler mode at approximately twice the pump wavevector~\cite{Stevenson:PRL2000,Baumberg:PRB2000,Krizhanovskii:PRL2006}. This process is illustrated in the bottom-left frame of Fig.\ref{fig:cavity_BEC}. Above some threshold value of the pump intensity, parametric oscillation sets in with a macroscopic occupation of signal and idler states. While the phase matching condition imposes that the sum of the signal and idler phases is locked to twice the pump phase, there is no restriction for the phase difference between signal and idler states. As a result, at the onset of parametric oscillation, a spontaneous U(1) symmetry breaking occurs associated to the emergence of extended spatial coherence~\cite{carusotto2005spontaneous,DelVallePRL2009}. This is precisely what we have defined as Bose Einstein condensation in Sec.\ref{sec:BEC}. 
In the following this scheme to create a polariton condensate will be referred to as ``OPO scheme''. From the theoretical point of view, this scheme has the important advantage of allowing for {\em ab initio} calculations, e.g. using truncated Wigner techniques~\cite{carusotto2005spontaneous,dagvadorj2015nonequilibrium}. 

Depending on the cavity parameters and on the precise excitation scheme, polariton condensates may present a thermal tail or an out of equilibrium population. For instance in the OPO scheme, the polariton distribution is not thermalized. For non-resonant excitation, a Bose-Einstein distribution of polaritons with a well defined temperature can be instead observed by photoluminescence experiments, provided the cavity lifetime is sufficiently long as compared to the thermalization time scales~\cite{sun2017bose}. 

In polariton systems, the relaxation and thermalization rates are strongly affected by the cavity-exciton spectral detuning. A phase diagram as a function of the excitation power and the detuning of the cavity mode from the exciton was calculated and experimentally explored in~\cite{KasprzakPRL2008,sun2017bose}. As it is illustrated in the bottom-center frame of Fig.\ref{fig:cavity_BEC}, a thermal Bose Einstein distribution is observed for zero or positive detunings, whereas a clearly out-of-equilibrium regime is found for negative detunings larger than the exciton-photon coupling (some authors indicate this configuration as ``polariton lasing"). 

Even though the concept of thermalization is often used in this polariton condensate literature to indicate the presence of a thermal tail in the population distribution among the modes, it is important to keep in mind that this is not an exhaustive proof of thermal equilibrium. As it was first proposed in~\cite{chiocchetta2017laser}, a rigorous evidence  could be obtained from fluctuation-dissipation relations to be assessed via more sophisticated measurements.

In contrast to the weak nonlinearity of photon condensate systems, the strong interactions between excitons are responsible for sizable inter-particle interactions within the polariton condensate and between the condensate and its environment with important experimental consequences. For instance, the interaction between a condensate and incoherent excitons induces a significant local blueshift of the polariton states. This enables to engineer interesting potential landscapes for polaritons by shining a laser beam with a suitable spatial profile on the cavity~\cite{Amo:PRB2010,Tosi:NatPhys2012,Gao2015, berloff2017realizing}. In this way, an all-optical transistor-like switch was realized in~\cite{gao2012polariton} by blocking the flow of a polariton condensate with optically-injected excitons.
Furthermore, a ring-shaped non-resonant excitation was shown in~\cite{Manni-PRL2011} to lead to a condensate at the center of the ring together with giant quantized vortices. Other routes for the generation of vortex/antivortex lattices via suitable spatial patternings of the incoherent pump spot have been demonstrated in~\cite{tosi2012}. 

A different strategy to generate ring-shaped configurations with a net angular momentum was pioneered in OPO polariton condensates and exploited to investigate fundamental features of polariton superfluidity~\cite{marchettivortexopo2010,Sanvitto:NatPhys2010,tosi2011onset}. In these works, condensation was forced to occur into the desired state by seeding the signal mode with a temporally short pulse with a spatial Laguerre-Gauss profile. The vortex imprinted into the signal mode (as well as the corresponding anti-vortex in the idler mode) was then shown to persist  for macroscopically long times. This observation provided a crucial experimental evidence of metastability of supercurrents in polariton condensates~\cite{Wouters:PRB2010}, in close analogy with the behaviour of standard equilibrium superfluids~\cite{Leggett:1999RMP}.
 
The important role of Coulomb exchange in the polariton–polariton interactions~\cite{Ciuti:PRB1998} results in a sizable spin-dependence of their strength: interaction between polaritons of same spin is always repulsive while interactions between anti-parallel polaritons can be weakly attractive~\cite{Vladimirova:PRB2010} and even display some Feshbach resonance effect across the biexciton energy~\cite{Carusotto:EPL2010,takemura2014polaritonic}. This spin-dependence of interactions is at the origin of rich bifurcation effects in the spin dynamics of a polariton condensate~\cite{Ohadi-PRX-2015}. 

Whereas most of the experimental works mentioned so far were performed at cryogenic temperatures, intense research efforts are being devoted to the development of novel materials displaying more robust excitonic resonances and enabling the exploration of polariton BEC and lasing at room temperature. Room temperature operation was first demonstrated in GaN based cavities~\cite{ChristopoulosPRL2007,Christmann2008}, then in ZnO~\cite{Lu2012,LiPRL2013}, organic materials~\cite{Kena-Cohen:NatPhot2010,Plumhof2014}, luminescent proteins~\cite{Dietriche2016}  and perovskites~\cite{Su2017,Su2020}. Realizing an electrically pumped polariton laser operating at room temperature is still a challenging open problem that is attracting a strong interest~\cite{Schneider2013,BattacharyaPRL2013,BhattacharyaPRL2014,Deveaud-CommentPhysRevLett2016,SuchomelPRL2018,Suchomel:APL2020}.
Another frontier is to observe BEC of polaritons emerging from the strong coupling between light and intersubband transitions in doped quantum wells~\cite{Dini:PRL2003,Anappara:PRB2009,Ciuti:PRB2005}. As compared to the exciton-polaritons discussed so far, such intersubband polaritons appear as very promising for the development of novel coherent light sources at longer wavelengths down to the Mid-Infrared and the THz regions~\cite{colombelli2015perspectives}. Further exciting research avenues are being opened by laser operation into phonon-polariton modes~\cite{Ohtani2019}.

\begin{table*}[]
    \centering \tiny
    \begin{tabular}{|c|c|c|c|c|c|}
    \hline
          {\bf System} & {\bf Mass} & {\bf Temperature} & {\bf Interactions} & {\bf Thermalization mechanism} & {\bf Properties} \\ \hline
          Cold Atoms~\cite{BECbook} & $\sim 10^5 m_e$ & nK & fully tunable & Collisions + Evaporation & Conservative, Thermal \\ \hline
          Laser & $\sim 10^{-5}\,m_e$ & not relevant & weak & none & Driven-Dissipative, Non-Equilibrium \\ \hline
          Photon BEC (Sec.) & $10^{-5}\,m_e$ & room & weak & absorption/emission by dye & Driven-Dissipative, Thermal$^*$ \\ \hline
          Exciton BEC & $m_e^*$ & $1$~K & Strong & Collisions, phonon scattering &  Driven-Dissipative, Thermal \\ \hline
          Polariton BEC & $10^{-5}\,m_e$ & 10-300~K & MF, strong & \makecell{Collisions, phonon scattering,\\ scattering on reservoir excitons} & Driven-Dissipative, Non-Equilibrium/Thermal$^*$ \\ \hline
          Classical condensation \\ of light waves & \multicolumn{2}{c|}{\parbox[c]{4cm}{not relevant \\ $t\leftrightarrow z$ mapping}}  & MF, moderate & Optical nonlinearities & Conservative, Time-Dependent \\ \hline
          \end{tabular}
    \caption{Comparative summary  of the characteristic parameters of different physical realizations of BEC: cold atoms, laser, photons, excitons, polaritons, classical propagating light. By ``Thermal$^{*}$'' we mean thermal distribution of population and not full thermalization according to the fluctuation-dissipation relations. By ``MF" we mean interactions at the mean-field level of GPE-like equations.}
    \label{tab:comparison}
\end{table*}

\subsection{Condensation of classical light waves in waveguides \label{sec:waveguide}}

A conceptually different example of photon condensation is provided by  condensation of classical light waves in cavity-less propagating geometries. 
In such systems, the role of time is played by the propagation direction $z$ along the waveguide: within the paraxial approximation, the propagation  of monochromatic light field can be recast in terms of a Gross-Pitaevskii equation (GPE) after the substitution $t\rightarrow z$. In contrast to cavity systems, no intrinsic radiative losses are present since there is no extra spatial direction in which photons can escape. Since also absorption losses are typically weak, the dynamics can be accurately described as a conservative one. The intensity-dependent refractive index of the medium provides effective photon-photon interactions that mediate thermalization of the classical waves. After first pioneering attempts in unbounded geometries~\cite{Sun:NatPhys2012,Santic:PRL2018}, thermalization and then condensation were found to be facilitated by confinement within multi-mode graded index waveguides~\cite{krupa2019multimode}, which provide an effective harmonic trapping potential to photons.

In a typical experiment, a monochromatic, yet spatially disordered classical light beam is injected into the waveguide and serves as the initial condition of the field. The amount of disorder in the incident field sets the total energy of the (typically non-thermal) initial state. 
If the optical nonlinearity is sufficiently strong and the propagation length sufficiently long, thermalization into a Rayleigh-Jeans-like distribution $n(\epsilon) = k_B T_{fin}/\epsilon$ is expected, where $\epsilon$ is the energy of the different transverse modes and the final temperature $T_{fin}$ is fixed by energy conservation~\cite{Connaughton:PRL2005,aschieri2011condensation,picozzi2014optical}. Conveniently, the UV blackbody catastrophe is ruled out in these experiments by the finite number of available modes in the waveguide. 

For a sufficiently low value of the energy of the incident beam, thermalization is associated to  condensation into the lowest waveguide mode, so that the transverse spatial profile of the beam gets rid of its speckle-like spatial modulations according to a self-cleaning phenomenon~\cite{krupa2017spatial,baudin2020rayleigh}. Examples of experimental images of such thermalization and condensation processes are shown in the bottom frames of Fig. \ref{fig:other_plat}. 
In these experiments, the transverse confinement was strong enough for the condensate to be stable against long-wavelength fluctuations. Experimental evidence of BKT features was instead reported in~\cite{situ2020dynamics}. 

As a final remark, it is important to emphasize that in all these experiments the incident light behaves a purely classical entity and preserves its monochromatic character while thermalizing to a Rayleigh-Jeans distribution. The full Bose-Einstein distribution would however be  recovered upon inclusion of quantum fluctuations, which give the beam a finite bandwidth~\cite{Larre:PRA2015,chiocchetta2016thermalization}. While the characteristic length scale of such quantum effects is typically excessively long in most condensation experiments so far, promising perspectives are opened by the realization of strongly interacting Rydberg polaritons in atomic clouds~\cite{Peyronel:Nature2012} and by recent observations of quantum phenomena in propagating condensates of light~\cite{steinhauer2021analogue}. 
In the context of this review, an intriguing long-term challenge is to demonstrate light condensation effects into a spatially clean and temporally quasi-monochromatic beam starting from fully incoherent light. As proposed in~\cite{chiocchetta2016thermalization},  such an observation may be facilitated by exploiting evaporation cooling techniques inspired by ultracold atomic gas experiments.

\begin{figure*}[htbp]
    \centering
    \includegraphics[width=\textwidth]{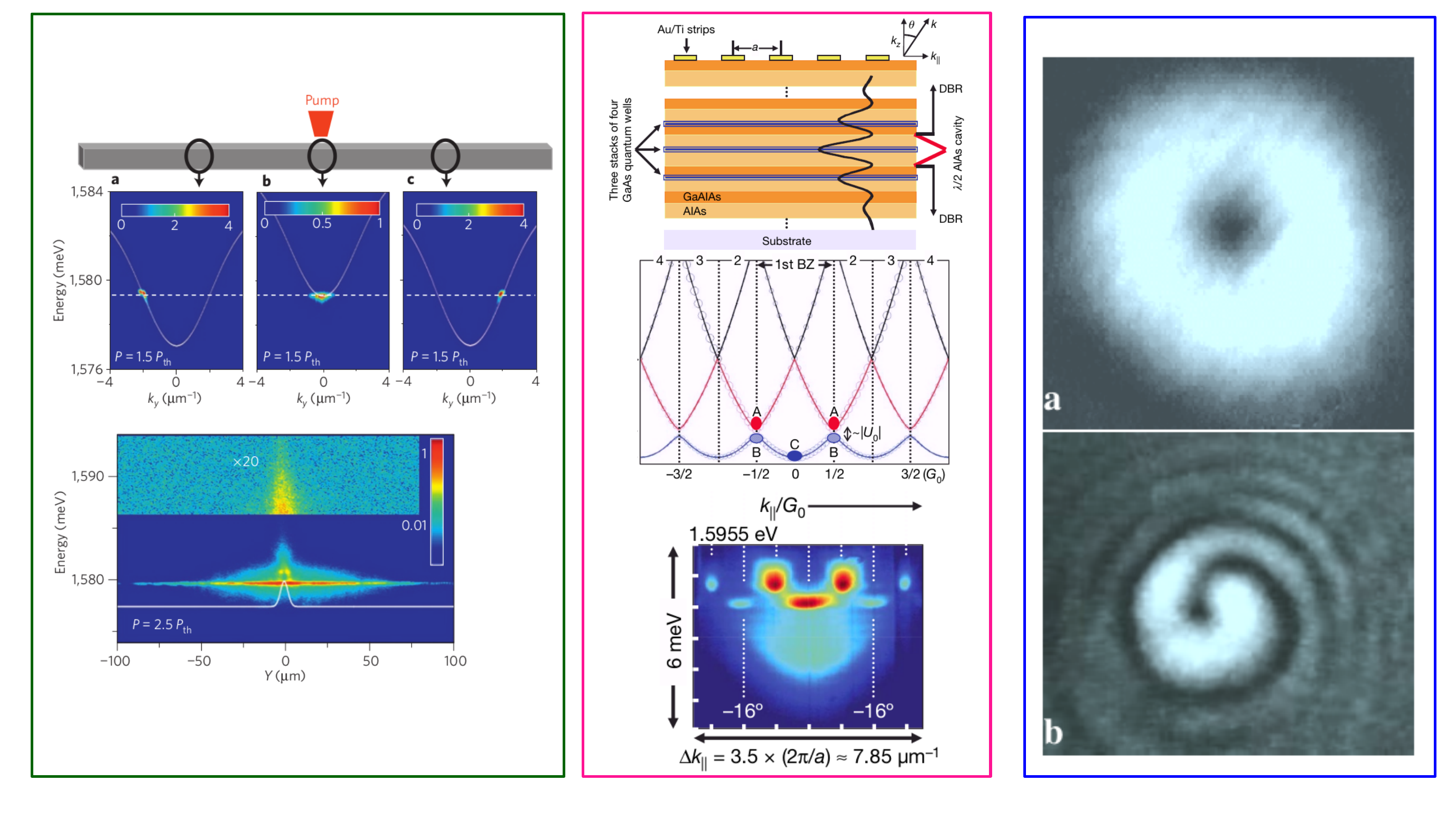}
    \caption{New features of non-equilibrium condensates. Left frame: ``polariton volcano" effect.  Sketch of the polariton wire and of the pumping geometry used in the experiment (top). Spatially-resolved energy-momentum distribution of the emission at three different points (center). Spatially- and energy-resolved distribution of the emission of the outward flowing condensate (bottom). Figure from Ref.\onlinecite{Wertz:NatPhys2010}.
    Center frame: condensation in an excited state.  Sketch of the polariton lattice used in the experiment (top). Polariton band dispersion (center). Energy-momentum distribution of the polariton emission showing peaks for condensation at $k=0,\pi$ (bottom). 
    Figure from Ref.\onlinecite{lai2007coherent}. Right frame: density and flow profile of a singly-charged quantized vortex in the coherent field of a photorefractive oscillator. Intensity (top) and interferogram displaying a spiral-shaped phase profile (bottom). Figure from Ref.\onlinecite{weiss1999solitons}.}
    \label{fig:non_equil}
\end{figure*}

\section{Non-equilibrium effects} 
\label{sec:non_eq}

While the appearance of a macroscopic coherence extending up to large spatial distances is the general signature of condensation, the driven-dissipative nature of the steady-state can lead to a much richer phenomenology as a result of non-Hermitian effects and broken time-reversal symmetry. The present Section is devoted to a review of some among the most intriguing such features.



Textbook thermodynamic arguments based on free energy minimization enforce that equilibrium BEC occurs into the lowest energy single-particle orbital~\cite{Huang}, which is possibly deformed by interactions~\cite{BECbook} but maintains a real-valued wavefunction in real space. On the other hand, the spatial shape of a non-equilibrium condensate is determined by a complex interplay of pumping, losses, kinetic energy and interactions and can display a variety of features. In particular, the complex-valued nature of the condensate wavefunction is responsible for breaking the $\mathbf{k}\leftrightarrow -\mathbf{k}$ symmetry of the far-field intensity distribution as discussed e.g. in~\cite{Wouters:PRB2008}.

The spontaneous generation of complex patterns in spatially extended lasers with large Fresnel numbers has been the subject of intense investigations for decades. Following early theoretical works~\cite{COULLET1989403,lugiato1990instabilities}, several experiments have shown vortex lattices in a variety of configurations, from photorefractive oscillators~\cite{Arecchi:PRL91,Mamaev_1996}, to VCSELs~\cite{Scheuer:Science1999} and end-pumped microchip lasers~\cite{Chen:PRA2001}. Quantized vortices in polariton condensates were first observed in~\cite{Lagoudakis:NatPhys2008}, soon followed by half-vortex structures involving the spin degrees of freedom~\cite{Lagoudakis:Science2009}; related vector vortices were reported in VCSEL devices with suitably designed feedback elements~\cite{Jimenez:PRL2017}. In all these systems, the appearance of the vortex lattice is the result of many different ingredients, from nonlinear mode-locking between quasi-degenerate modes to complex flows in the underlying disorder potential to polarization-dependent optical elements, with the common feature that it only appears when the coherent field occupies a large enough region in space.

An unexpected behaviour as a function of the pump spot size was observed in early polariton condensation experiments~\cite{Richard:PRL2005,Richard:PRB2005}. While usual condensation into the $\mathbf{k}\approx0$ states was recovered for relatively large spots; an extended, ring-shaped $\mathbf{k}$-space distribution  was observed for smaller spots, while preserving macroscopic coherence between the different $\mathbf{k}$-modes. As it was explained in~\cite{Wouters:PRB2008} and then experimentally confirmed in~\cite{Wertz:NatPhys2010}, this behaviour is a direct consequence of the relatively strong repulsive interactions between polaritons and between the polaritons and the excitonic reservoir, which induce a stationary radial current via a ``polariton volcano'' effect as illustrated in the left frame of  Fig.\ref{fig:non_equil}.


A further class of excited-state condensation phenomena occur when polariton condensates are subject to a periodic potential (center frame of Fig.~\ref{fig:non_equil})~\cite{lai2007coherent}. Depending on the system parameters and the pumping regime, condensation can occur into the $k=0$ lowest-energy state and/or at the top of the lowest band, at quasi-momentum $k=\pi$. 
Further insight in this physics was provided by the experiments in~\cite{Tanese:NatComm2013}, which showed polariton condensation into gap soliton states energetically located above the lowest band and bound by the positive interaction energy. 
Beyond this, 
polariton lattices \cite{schneider2016exciton} offer a  versatile platform to induce polariton condensation into a variety of other  excited states like d-orbital states~\cite{Kim-NaturePhys2011} or plaquette-states in a flat band~\cite{baboux.PhysRevLett.116.066402}. The spontaneous appearance of  vortex lattices with a corresponding breaking of time-reversal symmetry was predicted in~\cite{lledo2021spontaneous} for polariton condensates in the lowest Landau level of a strained honeycomb lattice.

While the long-range coherence of VCSELs and wide condensates in the lowest energy states is often spoiled by different dynamical instability mechanisms~\cite{HegartyPRL1999,bobrovska2014stability,bobrovska2018dynamical},
a robust long-distance coherence was experimentally reported for condensation in negative-mass states of polariton lattices around the maximum of an energy band~\cite{baboux2018unstable}. As we shall see below, this opens exciting experimental perspectives for the study of the critical properties of the non-equilibrium phase transition. 

Another more sophisticated strategy to stabilize a long-range coherent emission is being pursued in devices where condensation occurs into the chiral edge states of a two-dimensional topological photonic system~\cite{bahari2017nonreciprocal,bandres2018topological,ozawa2019topological}. In such {\em topolaser} devices, the mutual coherence between spatially separated regions is in fact maintained by the chiral motion of light around the edge of the system, preventing multi-mode laser operation into multiple localized modes even in the presence of sizable disorder~\cite{harari2018topological,secli2019theory}. A theoretical study of the coherence properties of the topolaser emission was reported in~\cite{amelio2020}, highlighting a rich nonlinear dynamics of phase fluctuations, the key role of the intrinsically periodic boundary conditions of edge modes, and the ensuing remarkable robustness of the coherence to disorder.

Whereas these phenomena are mostly dominated by 
conservative propagation and interaction effects, intriguing relaxation effects towards the thermodynamic equilibrium states were observed in other polariton experiments and theoretically described by extended forms of the GPE~\cite{Savona:PRB2010}. 
Condensation into the ground state of the optical trap potential that is naturally generated around the center of a hollow ring-shaped pump beam was observed in~\cite{askitopoulos2013polariton}.  The interplay between the fast ballistic expansion of high-energy incoherent polaritons and an efficient energy relaxation into low-velocity states was at the heart of the wide polariton condensates at rest observed in~\cite{caputo2018topological}.
Alternative strategies for inducing condensation into a specific mode have  been investigated by exploiting, e.g., the polarization of the excitation beam: in ring-like chains of coupled micropillar cavities, the orbital angular momentum of the polariton condensate turns out~\cite{CarlonZambon:19} to be aligned with the circular polarization of the excitation laser by spin-orbit coupling effects~\cite{Sala:PRX2015}.

Besides polariton systems, a rich mode-selection phenomenology was observed also in photon condensates. Here, the thermalization rate can be tuned by varying the detuning of the cavity mode with respect to the dye absorption/emission spectrum. This leads to markedly different behaviours  when the excitation is located away from the center of the parabolic trap~\cite{schmitt.PhysRevA.92.011602}: For a fast thermalization rate, the photons are able to relax and condense into the zero momentum state around the potential minimum. For a slow thermalisation rate, the interconversion of the initial potential energy at the excitation location into kinetic energy and back can lead to net flows and even self-oscillation behaviours analogous to mode-locked lasers.





Interesting consequences of pumping and dissipation have been investigated in the structure and dynamics of quantized vortices in non-equilibrium BECs. The dip in the condensate density at the vortex core leads in fact to a local reduction of particle losses and, in turn, to a localized excess gain. To restore the balance between driving and dissipation, an outward flow must then appear in the condensate, reflected in a spiral-shaped phase profile. Such spiral phase profiles are well known for the complex Ginzburg Landau equation of pattern formation theory~\cite{Aranson:RMP2002} and have been observed in photorefractive oscillators~\cite{weiss1999solitons}, as shown in the right frame of Fig.\ref{fig:non_equil}.

The presence of this outward flow has been theoretically predicted to have  profound consequences on both the microscopic vortex dynamics and the macroscopic features of the condensation phase transition. At the former level, the interaction energy between vortices and antivortices can turn repulsive and the vortex motion can display self-acceleration behaviours~\cite{gladilin2017interaction,Staliunas_book}.
At the latter level, the fact that the outward flows impede the recombination of vortices and antivortices has been highlighted as a mechanism destabilizing the quasi-long-range order of the non-equilibrium Berezinskii-Kosterlitz-Thouless (BKT) phase in very large two-dimensional systems~\cite{wachtel2016electrodynamic,gladilin2019noise-induced} and has been predicted to strongly affect the phase healing kinetics \cite{Gladilin2019multivortex}. 

\begin{figure*}[htbp]
    \centering
    \includegraphics[width=\textwidth]{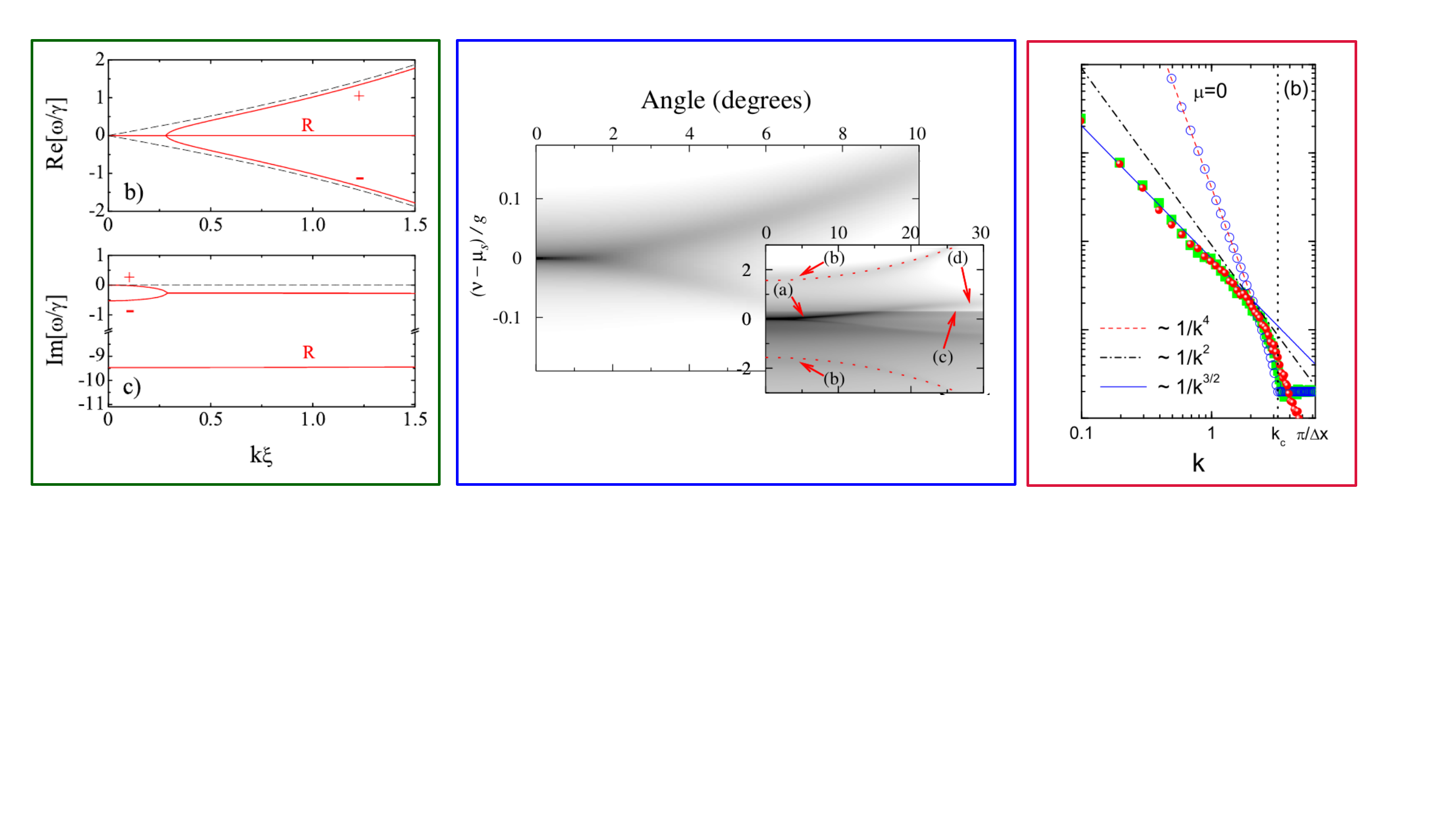}
    \caption{Left frame: Theoretical predictions for the real (top) and imaginary (bottom) parts of the Bogoliubov dispersion of a non-equilibrium condensate. Figure from Ref.\onlinecite{Wouters:PRL2007}. Center frame: prediction of a linearized theory for the angle- and energy-resolved photoluminescence of a non-equilibrium condensate displaying the diffusive feature at small $k$. The inset shows an enlarged view. Figure from Ref.\onlinecite{Szymanska:PRL2006}. 
    These two frames highlight the diffusive nature of the Goldstone mode of non-equilibrium BECs.
    Right frame: long-wavelength $k^{3/2}$ scaling of the decay time of the collective excitation mode, as numerically calculated with a fully nonlinear generalized GPE equation. Figure from Ref.\onlinecite{ji2015temporal}.}
    \label{fig:KPZ}
\end{figure*}

Further insight on the consequences of driving and dissipation on the phase transition was obtained by studies of the collective excitation spectrum of condensates. Where the Goldstone theorem still guarantees the presence of a soft branch whose dispersion tends to zero in the long wave length limit, the low-energy part of the spectrum no longer consists of hydrodynamic sound waves as in equilibrium condensates with $\omega_k\simeq c_s k$~\cite{BECbook} but displays a diffusive behavior $\omega_k\simeq - i \alpha k^2$~\cite{Szymanska:PRL2006,Wouters:PRL2007}. This behaviour is illustrated in the left and central frames of Fig.\ref{fig:KPZ}.

Quite remarkably, and in contrast to a naive application of the Landau criterion, superfluidity of non-equilibrium condensates was shown to largely survive this modification~\cite{Wouters:PRL2010,Keeling:PRL2011}. 
Following the lines of the theoretical proposal in~\cite{wouters2007goldstone}, first attempts to experimentally measure the diffusive nature of collective excitations were carried out in a spatially extended OPO system~\cite{Ballarini:PRL2009}. Measurements of sonic dispersion features of polariton BECs under incoherent pumping have been reported in~\cite{Utsonomiya:NatPhys2008,Ballarini:NatComm2020}. Pioneering studies of the dynamics of photon BECs were reported in~\cite{Walker:PRL2019}.

Within a linearized Bogoliubov description of collective excitations, the coherence properties of the driven-dissipative system turn out to be largely similar to those at thermal equilibrium. The fluctuations that are naturally associated to dissipation lead in fact to an effective temperature for the long wave length degrees of freedom \cite{Wouters:PRB2006,chiocchetta2013non}. Except for some peculiar features signalling the non-equilibrium condition, a similar overall agreement was also found for the BKT physics of two-dimensional geometries~\cite{dagvadorj2015nonequilibrium}.

Shortly after initial ground breaking works on polariton condensates, it was realized that the magnitude of long-wavelength fluctuations of the gradient of the condensate phase is so large that the linearized Bogoliubov approximation inevitably breaks down in the infrared limit for spatially large systems and nonlinear terms have to be included in the theoretical description.
As a result, the effective dynamics of the phase degrees of freedom falls in the so-called Kardar-Parisi-Zhang (KPZ) universality class \cite{kulkarni2013finite,altman2015two,wachtel2016electrodynamic,gladilin2014spatial,He:PRB2015,ji2015temporal,squizzato2018kardar,zamora2017tuning,he2017space} originally introduced in the context of interface roughening phenomena in crystal growth~\cite{kardar1986dynamic}. In the photonic BEC case, the KPZ physics is revealed in the spatio-temporal decay of the coherence of the emission (right frame of Fig.\ref{fig:KPZ}). In 1D, this is expected to display a peculiar scaling law $g^{(1)}(x,t) \sim \exp[-B |t|^{2\chi/z}\,f(t/x^{1/z}) ]$ in terms of an exactly known universal function $f$~\cite{prahofer2004exact}. This functional form results in a stretched-exponential form of the coherence decay along both the spatial and temporal directions, the exponents of $g^{(1)}(x,t=0) \sim \exp[-B |x|^{2\chi}]$ and $g^{(1)}(x=0,t) \sim \exp[-B |t|^{2\chi/z}]$ being respectively $\chi=1/2$ and $z=3/2$. In particular the temporal decay of $g^{(1)}(x=0,t)$ makes a clear difference from the $z=2$ exponent of the linearized Bogoliubov theory. For stronger noise levels, a more sophisticated phase transition toward a space-time vortex turbulent phase, directly related to the compactness of the phase variable, was predicted in Ref.~\cite{he2017space}.

Experimental evidence of 1D KPZ physics in polariton systems was recently reported in~\cite{Fontaine21} exploiting the robust negative mass polariton quasi-condensates that are produced in periodic potentials~\cite{baboux2018unstable}. Extension of these experiments to the 2D case is made challenging by the very long spatial and temporal scales involved in the phase dynamics. Still, it is of utmost interest because of the intriguing interplay of KPZ physics with those topological excitations, e.g. vortices, that are associated to the compact nature of the condensate phase variable~\cite{wachtel2016electrodynamic,zamora2017tuning}.


\section{Conclusions and future perspectives}
\label{sec:conclu}

\begin{figure*}[htbp]
    \centering
    \includegraphics[width=\textwidth]{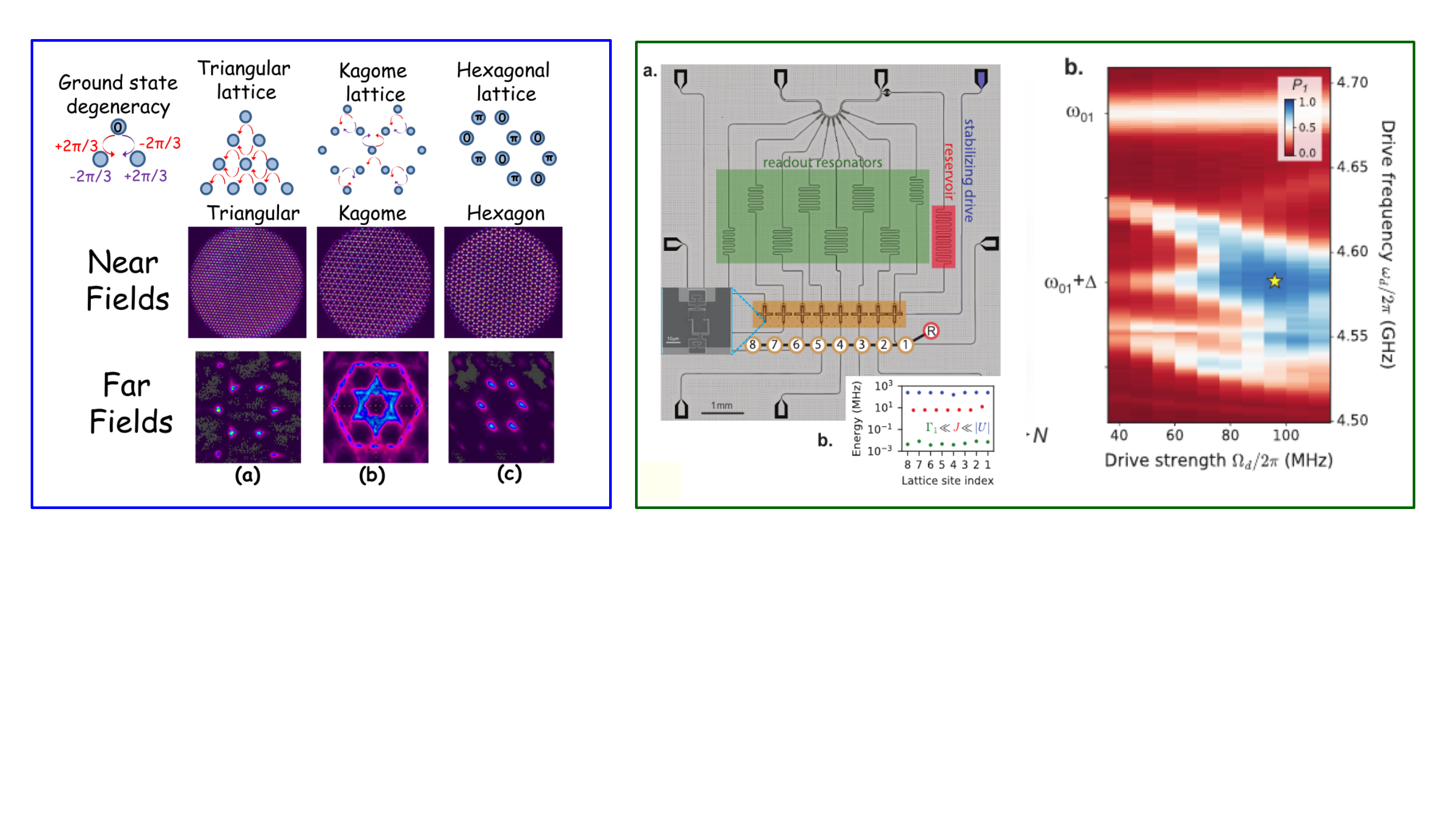}
    \caption{Future perspectives. Left frame: application of an array of coupled lasers to simulate a XY model in different lattice geometries (top). Near-field (center) and far-field (bottom) images of the laser field. Figure from Ref.\onlinecite{Davidson_laser2013}. Right frame: Mott insulator of strongly interacting photons. Left: Experimental circuit-QED set-up realizing a one-dimensional lattice of 8 sites. Right: Preparation fidelity of the target one-photon Mott-insulator state as a function of pumping parameters. Figure from Ref.\onlinecite{ma2019dissipatively}. }
    \label{fig:persp}
\end{figure*}

The rich phenomenology of non-equilibrium Bose-Einstein condensation reviewed in the previous Sections already hints at the rich exciting perspectives that these discoveries are opening to fundamental as well as applied research~\cite{colombelli2015perspectives,Sanvitto2016,fraser2016physics,BallariniDeLiberato,SegevBandres2021, colombelli2015perspectives}.

The most direct application of condensation effects is of course the realization of novel sources of coherent light.  These sources are expected to offer reduced threshold intensity below the one of standard lasers~\cite{malpuech2003polariton,liew2011polaritonic,fraser2016physics}, because of the slower decay rate of the reservoir pumping the BEC (excitons for polaritons, or dye excitations for photon BECs) as compared to electron/hole excitations, and of the reduced value of the required excitation density well below population inversion. Beyond providing a low energy consumption device, this low lasing threshold may strongly reduce heating effects and thus enable extending the range of wavelengths accessible to lasing operation. The advantages of polariton condensates will be technologically even more significant in devices where polaritons are robust up to room temperature~\cite{ChristopoulosPRL2007,Christmann2008,Lu2012,LiPRL2013,Plumhof2014,Kena-Cohen:NatPhot2010,Dietriche2016,Su2017,Su2020} and are electrically injected~\cite{Schneider2013,BattacharyaPRL2013,BhattacharyaPRL2014,Deveaud-CommentPhysRevLett2016, SuchomelPRL2018,Suchomel:APL2020}. Some of the platforms operating at room temperature are also highly relevant for reduced-cost technology~\cite{Kena-Cohen:NatPhot2010,Dietriche2016,Su2017,Su2020}. For  wavelength regions such as the Mid- and Far-Infrared where quantum cascade lasers so far suffer from temperature limitations~\cite{Khalatpour2021,Sirtori2021}, inter-subband polaritons have been proposed as an alternative avenue for room temperature operation~\cite{colombelli2015perspectives}. From a different perspective, polariton lasers in guiding geometry~\cite{Jamadi2018} have been proposed as  a relevant building block for integrated polariton circuits~\cite{Espinosa-Ortega2013Complete,Sanvitto2016,Rozas2020,BeierleinPRL2021}.

From a fundamental point of view, condensates of light provide a flexible platform to investigate open questions in non-equilibrium statistical mechanics. In this context, a special attention goes to models in the Kardar-Parisi-Zhang (KPZ) universality class~\cite{kardar1986dynamic} which are involved in a number of natural phenomena, from surface growth to flame propagation. In its compact version, KPZ physics was associated to subtle features of polariton quasi-condensation in reduced dimensions~\cite{altman2015two,wachtel2016electrodynamic,gladilin2014spatial,He:PRB2015,ji2015temporal,squizzato2018kardar,zamora2017tuning,he2017space}. First experimental evidence of KPZ physics in the spatio-temporal coherence properties of one-dimensional polariton condensates was observed by the authors in~\cite{Fontaine21}, opening the way towards more challenging investigations of KPZ physics in two-dimensions and of the robustness of coherence to disorder in topological laser devices~\cite{amelio2020}. Beyond these steady-state properties, a number of theoretical predictions are still awaiting experimental investigation, in particular on the temporal growth of coherence after a fast jump in the pump intensity~\cite{comaron2018dynamical} or during a slow Kibble-Zurek-style ramp~\cite{Matuszewski:PRB2014,Solnyshkov:PRL2016,Kulczykowski2017phase,Zamora:PRL2020} and on the meaning of superfluidity for non-equilibrium condensates~\cite{Wouters:PRL2010,Keeling:PRL2011}. 


As another promising development, condensation in complex geometries was proposed as a computational tool for the ground state of XY models with arbitrary coupling constants (left frame of Fig.\ref{fig:persp}). The mechanism is based on the gain competition idea, which favours accumulation of photons into the mode for which gain is the strongest. First demonstrations of this idea were  experimentally reported for arrays of coupled gas lasers~\cite{Davidson_laser2013}, polariton lattices~\cite{berloff2017realizing} as well as networks of coupled optical parametric oscillators~\cite{mcmahon2016fully}.


All these directions of research are presently attracting the interest of a growing community of researchers coming from different backgrounds and are putting optical sciences at the center of an interdisciplinary effort ranging from semiconductor optics, to quantum many-body physics, to non-equilibrium statistical mechanics, to computational sciences. 

On a longer run, a further new twist is expected to stem from the on-going advances in the development of optical devices with ultra-strong optical nonlinearities at the single-photon level, where single quanta of energy are able to substantially modify the refractive properties of the medium~\cite{imamoglu1997strongly,Verger:PRB2006,Chang:NatPhot2014}. In this regime, a variety of phase transitions with more complex order parameters are expected to take place and lead to exotic states of strongly correlated photonic matter displaying rich entanglement structures, such as at the superfluid to Mott insulator transition~\cite{Hartmann:NatPhys2006,Greentree:2006,Angelakis:PRA2007,Biella:PRA2017,Lebreuilly:PRA2017}, fermionized gases of impenetrable photons~\cite{Carusotto:PRL2009}, quantum magnetic models~\cite{rota17,rota2019quantum}, fractional quantum Hall fluids~\cite{Umucalilar:2012PRL,Kapit:2014PRX}. First experimental steps have demonstrated Mott insulators~\cite{ma2019dissipatively} (right frame of Fig.\ref{fig:persp}) and strong synthetic magnetic fields~\cite{Roushan:2016NatPhys} in circuit-QED platforms, small Laughlin droplets in Rydberg polariton systems~\cite{clark2020observation}, and many more investigations are presently in progress~\cite{carusotto2020photonic}.

Looking at what happened in the last decades with lasers and is presently happening with the other forms of photon condensation phenomena reviewed in this work, we can safely anticipate that the realization and manipulation of the quantum entanglement inherent in these new states of photonic matter will be a fundamental resource for a number of optical devices for quantum technology tasks.

\begin{acknowledgments}
JB acknowleges financial support from the Paris Ile-de-France R\'egion in the framework of DIM SIRTEQ,  the EU project "QUANTOPOL" (846353), the H2020-FETFLAG project PhoQus (820392), the QUANTERA project Interpol (ANR-QUAN-0003-05), the French National Research Agency project Quantum Fluids of Light (ANR-16-CE30-0021), and the French RENATECH network.
IC acknowledges financial support from the European Union FET-Open grant “MIR-BOSE” (n. 737017),
from the H2020-FETFLAG-2018-2020 project ”PhoQuS” (n.820392), from the Provincia Autonoma di
Trento, from the Q@TN initiative, and from Google via the quantum NISQ award.
MW acknowledges financial support from the FWO-Vlaanderen (grant nr. G016219N).
\end{acknowledgments}
\bibliography{RMPlight_v24,Bibliography_Iacopo,Bibliography_Michiel,Bibliography_Common,more_biblio.bib}

\end{document}